# Early Prediction of Mortality in Critical Care Setting in Sepsis Patients Using Structured Features and Unstructured Clinical Notes

Jiyoung Shin
*Dept. of Preventive Medicine*
*Feinberg School of Medicine*
*Northwestern University*
Chicago, IL, U.S.A
jiyoung.shin@northwestern.edu

Yikuan Li
*Dept. of Preventive Medicine*
*Feinberg School of Medicine*
*Northwestern University*
Chicago, IL, U.S.A
yikuan.li@northwestern.edu

Yuan Luo (corresponding)
*Dept. of Preventive Medicine*
*Feinberg School of Medicine*
*Northwestern University*
Chicago, IL, U.S.A
yuan.luo@northwestern.edu

***Abstract*— Sepsis is an important cause of mortality, especially in intensive care unit (ICU) patients. Developing novel methods to identify early mortality is critical for improving survival outcomes in sepsis patients. Using the MIMIC-III database, we integrated demographic data, physiological measurements and clinical notes. We built and applied several machine learning models to predict the risk of hospital mortality and 30-day mortality in sepsis patients. From the clinical notes, we generated clinically meaningful word representations and embeddings. Supervised learning classifiers and a deep learning architecture were used to construct prediction models. The configurations that utilized both structured and unstructured clinical features yielded competitive F-measure of 0.512. Our results showed that the approaches integrating both structured and unstructured clinical features can be effectively applied to assist clinicians in identifying the risk of mortality in sepsis patients upon admission to the ICU.**

## I. Introduction

Sepsis is a life-threatening organ dysfunction and a major public health issue. It is a common and economically important disease leading to 5.3 million death annually. The estimated overall mortality of sepsis patients is 30% [1-3]. Recently, sepsis was defined as a "life-threatening organ dysfunction caused by a dysregulated host response to infection" by The European Society of Intensive Care Medicine/Society of Critical Care Medicine Third International Consensus Definitions for Sepsis and Septic Shock task force (The Sepsis-3 task force) [2].

Early diagnosis and identification of sepsis are important to evaluate the patients' status and improve their survival outcomes. Furthermore, because of the vague definitions of sepsis syndrome, unknown infection sources, and higher risk of mortality, developing an effective and reliable prognostic prediction model for sepsis patients is important. Such models could help to predict the prognosis of the patients more efficiently, inform the allocation of public health resources, and support clinical decision-making.

Due to the increasing usage of electronic health records (EHRs), it is becoming easier to access comprehensive and extensive clinical data to predict health outcomes in the population. There are also many previous studies predicting mortality in patients [4-7]. However, many studies have largely focused on predicting mortality in specific population, such as elderly or patients who had cardiovascular surgery. These approaches could also draw meaningful conclusions in specific population, but it is also needed to build a mortality prediction model in general sepsis patients. Furthermore, many previous studies used several algorithm methods but did not include various laboratory information that is known to be effective for predicting the disease [8]. Finally, many of the mortality prediction studies have been conducted based solely on structured EHR data and did not incorporate unstructured clinical notes which could be ubiquitously utilized in other medical institutions as well [9].

Therefore, in this study we included a wide array of predictors including demographic characteristics and various physiological factors from the laboratory test that are recorded in the EHR, to predict mortality in sepsis-3 patients. Moreover, we used structural data (e.g., physiological variables) as well as unstructured intensive care unit (ICU) clinical notes data to construct the prediction model. These clinical notes are written by many clinical experts, including physicians and nurses, and could provide a comprehensive picture of patients' pathological statuses and aid in the development of a powerful model to predict the mortality in sepsis patients.

## II. Methods

### *A. Database and Study Population*

Data for this study were acquired from Medical Information Mart for Intensive Care III (MIMIC-III). MIMIC-III captures de-identified health information for more than 46,000 patients admitted to the critical care units at Beth Israel Medical Center between 2001 and 2012 [10]. We retrospectively defined the cohort for this study as patients meeting the criteria for sepsis-3, which is the new criteria evaluating sepsis: score equal or greater than 2 in Sequential Organ Failure Assessment (SOFA), in a defined context of suspected infection [2, 11]. The participants were admitted from 2008 to 2012 as in previous research. This is because of the accessibility of antibiotic prescriptions and explicit sepsis codes. Furthermore, the group of admissions were easily identifiable between 2008 to 2012 [12]. The initial ICU admission of patients was used to only focus on the first admission of patients with multiple admissions. The following exclusion criteria [12] were applied: 1) age less than 16 years,

2) previous cardiac surgery experience, 3) suspected infection more than 24 hours before and after the admission to the ICU.

### *B. Demographic and Physiological Structured Features*

We developed an SQL script to extract the patients and various characteristics of the patients, including demographic factors, physiological measurements, and clinical notes, from the database [13]. For the demographic characteristics, we extracted patients' age, gender, ethnicity, SOFA score assessed during their first ICU day, presence of metastatic cancer or diabetes, and so on. We included these factors in the structured model since these demographic factors might affect to the incidence of sepsis and mortality related to infection [14]. For the physiological measurements, we selected potential physiological variables related to mortality and extracted the first measurements of the patients after their ICU admission. Using the first measurements to build prediction model could be helpful to alert clinicians to predict the high risk of mortality soon after ICU admission. A total of 31 physiological items and 13 demographic characteristics were selected, and 44 items were selected as final structured features in the analysis (Table I). We excluded the physiological measurement outliers using the previously established reasonable range of measurements provided by the source code of Harutyunyan et al [15]. There were no missing values in the demographic characteristics except for body mass index (BMI), so we performed multiple imputations of physiological measurements and BMI using the "mice" package in R software version 4.0.5 [16]. Except the BMI (48%, proportion of missing values), aspartate aminotransferase (41%), base excess (35%), blood total bilirubin (39%), carbon dioxide (35%), international normalized ratio (11%), lactate (33%), ph (33%), partial thromboplastin time (11%), and albumin (48%), other physiological measurements had missing values of less than 2% before using the imputation method.

### *C. Natural Language Processing of the Unstructured Clinical Notes*

We extracted the clinical notes data during the first day of ICU admission and excluded the notes from the discharge summary category. After excluding patients who did not have any notes record, 5,396 patients were included in the final analysis. To handle and interpret the clinical notes, we first performed preprocessing steps. Masked protected health information (PHI) was removed from the clinical notes, and then the notes were converted into structured features for use with the machine learning classifier.

We utilized the unigram bag-of-words model to convert the unstructured notes to normalized lexical variants. Following NCBI guideline, 313 stop words are applied, and features with less than a 10-document-frequency were removed to reduce the noise. We also applied term frequency-inverse document frequency (tf-idf) weighting adjustment [17]. After the pre-processing steps, 7248 words were kept in the dictionary of the text corpus, and 7248 bag-of-words vectors were created for each patient's clinical notes. We used these features for the unstructured features model, but we also concatenated those vector features with structured features and created combined feature vectors.

TABLE I. UNIVARIATE CHARACTERISTICS FOR DEMOGRAPHIC AND PHYSIOLOGICAL CHARACTERISTICS (N=5,396)

| Variable | Value | Variable | Value | Variable | Value |
|---|---|---|---|---|---|
| Age, years | 65.5±17.6 | Race | | Chloride (mEq/L) | 105±6.85 |
| Sex | | White | 392 (72.7) | Carbon dioxide (mEq/L) | 24.6±5.77 |
| Male | 2382 (44.1) | Black | 473 (8.77) | Diastolic blood pressure (mmHg) | 66.7±17.5 |
| Female | 3014 (55.9) | Hispanic | 182 (3.37) | Glasgow coma scale motor | 4.92±1.80 |
| Metastatic cancer | | Asian | 167 (3.09) | Glucose (mg/dL) | 148±72.1 |
| Yes | 311 (5.76) | Other | 654 (12.1) | Heart rate (bpm) | 91.1±20.4 |
| No | 5085 (94.2) | Marital status | | Hematocrit (%) | 32.3±6.17 |
| Diabetes | | Divorce | 328 (6.08) | Hemoglobin (g/dL) | 10.8±2.09 |
| Yes | 1530 (28.4) | Married | 2379 (44.1) | International normalized ratio | 1.50±0.75 |
| No | 3866 (71.7) | Single | 1536 (28.5) | Lactate (mmol/L) | 2.16±1.74 |
| BMI | 28.6±8.55 | Widowed | 795 (14.7) | Magnesium (mg/dL) | 1.92±0.44 |
| Elixhauser score | 3.80±7.00 | Unknown | 358 (6.63) | Mean arterial blood pressure (mmHg) | 82.0±18.5 |
| Mechanical Ventilation | | Admission type | | Ph (unit) | 7.36±0.10 |
| Yes | 2586 (47.9) | Elective | 310 (5.74) | Platelet count (K/uL) | 214±116 |
| No | 2810 (52.1) | Emergency | 5026 (93.1) | Partial thromboplastin time (sec) | 36.8±21.7 |
| SOFA | 4.61±3.10 | Urgent | 60 (1.11) | Red blood cell count (m/uL) | 3.58±0.71 |
| SIRS | 2.92±0.93 | Aspartate aminotransferase (IU/L) | 191±837 | Respiration rate (insp/min) | 19.4±6.64 |
| Insurance type | | Base excess (mEq/L) | -1.52±5.45 | Blood oxygen saturation (%) | 96.7±4.47 |
| Government | 157 (2.91) | Bicarbonate (mEq/L) | 22.9±4.95 | Systolic blood pressure (mmHg) | 124±25.1 |
| Medicaid | 532 (9.86) | Blood serum creatinine(mg/dL) | 1.55±1.64 | Temperature (Celsius) | 36.6±1.03 |
| Medicare | 3123 (57.9) | Blood serum potassium (mEq/L) | 4.18±0.79 | Urine output (ml) | 214±207 |
| Private | 1538 (28.5) | Blood sodium (mEq/L) | 138±5.66 | White blood cell count (K/uL) | 12.7±12.7 |
| Self-pay | 46 (0.85) | Blood total bilirubin (mg/dL) | 1.56±3.38 | Blood albumin (g/dL) | 3.01±0.61 |
| | | Blood urea nitrogen (mg/dL) | 29.0±24.1 | | |

Data are presented as mean±SD, number (%)

Abbreviations: SD, standard deviation; BMI, body mass index; SOFA, sequential organ failure assessment; SIRS, systemic inflammatory response syndrome

### D. *Machine Learning Classifiers and Convolutional Neural Networks*

We built and applied various machine learning classifiers, including L1- and L2- regularized logistic regression (LR), random forest (RF), L1- and L2- regularized support vector machine (SVM) with linear kernel [18], XGBoost and multi-layer perceptron (MLP). Machine learning classifiers were generated using the Scikit-learn packages of Python to find the best prediction model [19].

We used two clinical outcomes for the prediction: hospital mortality and 30-day mortality of the sepsis-3 patients. Overall data sets were split into training and test set, and the split ratio was 7:3. The hospital mortality rate in the overall dataset was 12.94%, while the 30-day mortality rate was 16.51% (Table II). Since our mortality outcomes are imbalanced, we used the 'class_weight' parameter to down weight the majority class and balance the outcomes in the LR, RF, SVM models. We also used 'imblearn' library to apply random under-sampling as 1:4 ratio and compared the results with the model that did not use under-sampling method and only used 'class_weight' parameter. To avoid the overfitting and tune the model parameters, we performed five-fold cross-validation on the training set, and the best parameters were applied to the test set to assess the prediction performance of each model.

We also used convolutional neural networks (CNN) to integrate unstructured features and structured features. First, three one-dimensional convolutional layers with different filter sizes were built on the pre-trained word embedding. Max-pooling was used to select the most important features with the highest value in each of the three convolutional feature maps. Then, the pooling results were concatenated with the structured features. The concatenated hidden features were then fed into two fully connected layers, each followed by a dropout and ReLU activation layer. Finally, softmax function was applied to yield the probability distribution of the mortality labels. The models were trained with a learning rate of 0.0005; the batch size was 32; and the max epoch number was set to 20 with early stopping conditioned on the validation split, which was 10% randomly sampled from the training split. The word-embedding was pretrained on MIMIC-III dataset. We also used random under-sampling method in the CNN training to address the class imbalance issue. CNN model which only used word features was also performed with the same parameters.

TABLE II. MORTALITY DISTRIBUTION OVERALL AND IN TRAINING AND TEST SETS.

| Outcome | Set | Total number | Yes | No | Mortality rate |
|---|---|---|---|---|---|
| Died in the hospital | Overall | 5396 | 698 | 4698 | 12.94% |
| | Training | 3777 | 483 | 3294 | 12.79% |
| | Test | 1619 | 215 | 1404 | 13.28% |
| Died within 30 days[a] | Overall | 5396 | 891 | 4505 | 16.51% |
| | Training | 3777 | 631 | 3146 | 16.71% |
| | Test | 1619 | 260 | 1359 | 16.06% |

[a]30-day mortality after ICU admission

To evaluate the model's performance, we reported the area under the receiver (AUC), precision, recall, and F1-measure score of each model predicting hospital mortality and 30-day mortality. Furthermore, a permutation test was applied and repeated 1,000 times to identify whether the AUC was significantly different between the best machine learning classifiers with different features. P-value less than 0.05 was considered significant.

## III. RESULTS

The performance of the supervised machine learning classifiers (LR, RF, SVM, XGBoost, MLP) and CNN models in predicting mortality in sepsis patients is reported in this section. Table III shows the results for each machine learning model using structured features only. The baseline predictor of each representation using the RF algorithm yielded an AUC in the range of 0.80 to 0.83 and an F-measure ranging from 0.46 to 0.48. The L1-regularized SVM model predicting 30-day mortality showed the best F-measure compared to the other models (AUC of 0.822, F-measure of 0.508), and the model predicting hospital mortality showed better AUC when compared to the same model predicting 30-day mortality.

TABLE III. MACHINE LEARNING MODEL RESULTS FOR EACH MACHINE LEARNING ALGORITHM USING STRUCTURED FEATURES ONLY

| Outcome | Under sampling | ALG[a] | AUC | P | R | F |
|---|---|---|---|---|---|---|
| Hospital mortality | N/A | L1-LR | 0.836 | 0.336 | 0.730 | 0.460 |
| | | L2-LR | 0.835 | 0.335 | 0.726 | 0.458 |
| | | RF | 0.824 | 0.364 | 0.633 | 0.462 |
| | | **L1-SVM** | **0.836** | **0.344** | **0.744** | **0.471** |
| | | L2-SVM | 0.833 | 0.259 | 0.800 | 0.391 |
| | | XGBOOST | 0.802 | 0.525 | 0.242 | 0.331 |
| | | MLP | 0.827 | 0.552 | 0.247 | 0.341 |
| | 1:4 | L1-LR | 0.836 | 0.344 | 0.726 | 0.466 |
| | | L2-LR | 0.835 | 0.342 | 0.726 | 0.465 |
| | | RF | 0.825 | 0.371 | 0.633 | 0.467 |
| | | **L1-SVM** | **0.835** | **0.351** | **0.730** | **0.474** |
| | | L2-SVM | 0.833 | 0.342 | 0.716 | 0.463 |
| | | XGBOOST | 0.807 | 0.457 | 0.349 | 0.396 |
| | | MLP | 0.810 | 0.497 | 0.405 | 0.446 |
| 30-day mortality | N/A | L1-LR | 0.825 | 0.367 | 0.742 | 0.491 |
| | | L2-LR | 0.825 | 0.370 | 0.746 | 0.494 |
| | | RF | 0.800 | 0.374 | 0.635 | 0.471 |
| | | **L1-SVM** | **0.821** | **0.380** | **0.742** | **0.503** |
| | | L2-SVM | 0.824 | 0.369 | 0.750 | 0.495 |
| | | XGBOOST | 0.805 | 0.528 | 0.323 | 0.401 |
| | | MLP | 0.775 | 0.547 | 0.292 | 0.381 |
| | 1:4 | L1-LR | 0.825 | 0.370 | 0.746 | 0.494 |
| | | L2-LR | 0.825 | 0.370 | 0.750 | 0.496 |
| | | RF | 0.808 | 0.414 | 0.562 | 0.476 |
| | | **L1-SVM** | **0.822** | **0.383** | **0.754** | **0.508** |
| | | L2-SVM | 0.825 | 0.368 | 0.750 | 0.494 |
| | | XGBOOST | 0.809 | 0.503 | 0.350 | 0.413 |
| | | MLP | 0.795 | 0.467 | 0.323 | 0.382 |

Abbreviations: ALG, algorithm; L1-/L2-, L1/L2 regularization; LR, Logistic Regression; RF, Random Forest; SVM, Support Vector Machine; MLR, multi-layer perceptron; P, Precision; R, Recall; F, F-measure;
[a]Best F-measure in each survival outcome and under-sampling model are marked in bold

Table IV shows the machine learning model results using unstructured clinical notes only, which used bag-of-words features. Models predicting hospital mortality generally showed better AUC than those predicting 30-day mortality. For example, the L2-regularized SVM model predicting 30-day mortality had an AUC of 0.747. However, the L2-regularized SVM model that did not use the under-sampling method showed the best performance in predicting hospital mortality (AUC of 0.760, F-measure of 0.413). Moreover, the model that only used unstructured clinical notes did not lead to better AUC or F-measure when compared to the models using only structured features.

Table V shows the model results using both structured features and unstructured clinical notes. The results indicate higher performance compared to the same LR and L2-SVM models shown in Table III and IV, which only used unstructured clinical notes or structured features. The L2-regularized LR model which did not use the under-sampling method and constructed using combined features yielded the highest F-measure in 30-day mortality prediction (AUC of 0.842, F-measure of 0.512).

TABLE IV. MACHINE LEARNING MODEL RESULTS FOR EACH MACHINE LEARNING ALGORITHM USING UNSTRUCTURED CLINICAL NOTES ONLY

| Outcome | Under sampling | ALG[a] | AUC | P | R | F[a] |
|---|---|---|---|---|---|---|
| Hospital mortality | N/A | L1-LR | 0.747 | 0.276 | 0.586 | 0.376 |
| | | L2-LR | 0.761 | 0.320 | 0.577 | 0.412 |
| | | RF | 0.728 | 0.284 | 0.460 | 0.351 |
| | | L1-SVM | 0.726 | 0.255 | 0.577 | 0.354 |
| | | **L2-SVM** | **0.760** | **0.320** | **0.581** | **0.413** |
| | | XGBOOST | 0.713 | 0.349 | 0.102 | 0.158 |
| | | CNN | 0.673 | 0.319 | 0.102 | 0.155 |
| | 1:4 | L1-LR | 0.741 | 0.277 | 0.628 | 0.385 |
| | | **L2-LR** | **0.759** | **0.299** | **0.591** | **0.397** |
| | | RF | 0.725 | 0.259 | 0.465 | 0.333 |
| | | L1-SVM | 0.735 | 0.261 | 0.498 | 0.342 |
| | | L2-SVM | 0.758 | 0.295 | 0.605 | 0.396 |
| | | XGBOOST | 0.677 | 0.387 | 0.214 | 0.275 |
| | | CNN | 0.690 | 0.275 | 0.284 | 0.279 |
| 30-day mortality | N/A | L1-LR | 0.739 | 0.304 | 0.615 | 0.407 |
| | | L2-LR | 0.746 | 0.317 | 0.592 | 0.413 |
| | | RF | 0.694 | 0.283 | 0.458 | 0.349 |
| | | L1-SVM | 0.695 | 0.271 | 0.573 | 0.368 |
| | | **L2-SVM** | **0.747** | **0.325** | **0.569** | **0.413** |
| | | XGBOOST | 0.696 | 0.463 | 0.146 | 0.222 |
| | | CNN | 0.679 | 0.500 | 0.077 | 0.133 |
| | 1:4 | L1-LR | 0.736 | 0.296 | 0.615 | 0.400 |
| | | L2-LR | 0.744 | 0.317 | 0.577 | 0.409 |
| | | RF | 0.689 | 0.279 | 0.450 | 0.344 |
| | | L1-SVM | 0.684 | 0.269 | 0.565 | 0.365 |
| | | **L2-SVM** | **0.744** | **0.315** | **0.588** | **0.411** |
| | | XGBOOST | 0.691 | 0.369 | 0.200 | 0.259 |
| | | CNN | 0.696 | 0.464 | 0.123 | 0.195 |

Abbreviations: ALG, algorithm; L1-/L2-, L1/L2 regularization; LR, Logistic Regression; RF, Random Forest; SVM, Support Vector Machine; CNN, convolutional neural networks; P, Precision; R, Recall; F, F-measure;
[a]Best F-measure in each survival outcome and under-sampling model are marked in bold

Our results showed that the random under-sampling method did not generally improve the performance of the LR, and SVM models, but the performance of XGBoost and CNN model did increase after using the random under-sampling method. The L2-LR classifiers in combined model, resulted in a competitive AUC score over 0.841. In addition, the recall score was higher than the precision score for the LR and SVM algorithms. Since we are trying to capture as many future mortality onsets in sepsis patients and thus could tolerate modest false alarm, this result is well suited to the clinical application of the mortality alarm. Regarding the CNN result, the under-sampled model predicting hospital mortality yielded the best F-measure score of 0.291. Although the performance of the CNN increased in the under-sampled model when compared to the model that did not use under-sampling method, all CNN model in this result did not outperform the best non-CNN classifiers despite hyperparameter tuning.

Table VI shows the permutation test results of the best models with different features. The AUC of the model that used structured features was significantly higher than that of the model that used unstructured features regardless of the sampling method and survival outcome. Similarly, the AUC of the model that used combined features was significantly higher than that of the model that used unstructured features. When comparing the structured features model and combined features model, the AUC of the best models predicting 30-day mortality which used under-sampling method differed significantly. This result could indicate that the performance of combined model is higher than that of structured model.

TABLE V. MACHINE LEARNING MODEL RESULTS FOR EACH MACHINE LEARNING ALGORITHM USING STRUCTURED FEATURES COMBINED WITH UNSTRUCTURED CLINICAL NOTES

| Outcome | Under sampling | ALG[a] | AUC | P | R | F |
|---|---|---|---|---|---|---|
| Hospital mortality | N/A | L1-LR | 0.847 | 0.359 | 0.721 | 0.479 |
| | | **L2-LR** | **0.853** | **0.384** | **0.693** | **0.494** |
| | | RF | 0.812 | 0.395 | 0.535 | 0.455 |
| | | L1-SVM | 0.847 | 0.352 | 0.744 | 0.478 |
| | | L2-SVM | 0.854 | 0.376 | 0.684 | 0.485 |
| | | XGBOOST | 0.799 | 0.612 | 0.242 | 0.347 |
| | | CNN | 0.739 | 0.500 | 0.098 | 0.163 |
| | 1:4 | **L1-LR** | **0.847** | **0.359** | **0.744** | **0.484** |
| | | L2-LR | 0.853 | 0.356 | 0.684 | 0.468 |
| | | RF | 0.810 | 0.354 | 0.605 | 0.447 |
| | | L1-SVM | 0.843 | 0.345 | 0.721 | 0.467 |
| | | L2-SVM | 0.853 | 0.359 | 0.684 | 0.471 |
| | | XGBOOST | 0.823 | 0.535 | 0.353 | 0.426 |
| | | CNN | 0.734 | 0.403 | 0.251 | 0.209 |
| 30-day mortality | N/A | L1-LR | 0.839 | 0.388 | 0.712 | 0.502 |
| | | **L2-LR** | **0.842** | **0.406** | **0.692** | **0.512** |
| | | RF | 0.790 | 0.389 | 0.565 | 0.461 |
| | | L1-SVM | 0.831 | 0.375 | 0.723 | 0.494 |
| | | L2-SVM | 0.842 | 0.406 | 0.677 | 0.507 |
| | | XGBOOST | 0.783 | 0.559 | 0.308 | 0.397 |
| | | CNN | 0.761 | 0.593 | 0.062 | 0.112 |
| | 1:4 | L1-LR | 0.839 | 0.392 | 0.723 | 0.508 |
| | | L2-LR | 0.841 | 0.402 | 0.700 | 0.511 |
| | | RF | 0.790 | 0.372 | 0.585 | 0.454 |
| | | L1-SVM | 0.829 | 0.376 | 0.735 | 0.497 |
| | | **L2-SVM** | **0.841** | **0.404** | **0.700** | **0.512** |
| | | XGBOOST | 0.800 | 0.526 | 0.346 | 0.418 |
| | | CNN | 0.723 | 0.531 | 0.200 | 0.291 |

Abbreviations: ALG, algorithm; L1-/L2-, L1/L2 regularization; LR, Logistic Regression; RF, Random Forest; SVM, Support Vector Machine; CNN, convolutional neural networks; P, Precision; R, Recall; F, F-measure;
[a]Best F-measure in each survival outcome and under-sampling model are marked in bold

## IV. DISCUSSION

In this study, we examined the advantages of using combined features in predicting mortality in sepsis patients by comparing the performance scores between the models. Our results showed that the L2-regularized LR model constructed using bag-of words and structured features yielded the highest F-measure in 30-day mortality prediction. A logistic regression model is a traditional but powerful statistic method and it is widely used to predict the mortality and other outcomes [20-22].

We further ranked this L2-regularized LR model's coefficient of structured features to identify important demographic and physiological factors predicting sepsis mortality. Table VII showed the top structural features associated with an increased 30-day mortality risk.

TABLE VI. PERMUTATION TEST RESULTS AMONG BEST MACHINE LEARNING MODEL USING DIFFERENT FEATURES

| Outcome | Under sampling | Str vs. Unstr | | Str vs. Str + Unstr | | Unstr vs. Str + Unstr | |
|---|---|---|---|---|---|---|---|
| | | Model | P-value | Model | P-value | Model | P-value |
| Hospital mortality | N/A | **L1-SVM** vs. L2-SVM | **<0.001** | L1-SVM vs. L2-LR | 0.338 | L2-SVM vs. **L2-LR** | **<0.001** |
| | 1:4 | **L1-SVM** vs. L2-LR | **0.003** | L1-SVM vs. L1-LR | 0.459 | L2-LR vs. **L1-LR** | **<0.001** |
| 30-day mortality | N/A | **L1-SVM** vs. L2-SVM | **<0.001** | L1-SVM vs. L2-LR | 0.211 | L2-SVM vs. **L2-LR** | **<0.001** |
| | 1:4 | **L1-SVM** vs. L2-SVM | **<0.001** | L1-SVM vs. **L2-SVM** | **0.010** | L2-SVM vs. **L2-SVM** | **<0.001** |

Best models marked in bold in Table III, Table IV, and Table V were compared
AUC score was compared to calculate observed difference and p-value between each model
Bolded p-values indicate same bolded model showed significantly higher AUC compared to the other model

Our top feature was metastatic cancer, which is an obvious primary cause of the mortality [23]. There is also as a research that many of cancer patients are treated by chemotherapy, and this treatment could increase the risk of infection and sepsis [24]. Mechanical ventilation and hemoglobin are also known risk factors of mortality and associated with patients with more severe illness or sepsis [25, 26]. Furthermore, blood albumin was one of the top 10 structured features in the model. Albumin is considered an important factor with regard to nutritional status and is also known to be related to inadequate dialysis, fluid overload, and infectious diseases [27-29]. Several studies have shown that serum albumin decrement is an independent predictor of mortality risk and could be useful for predicting mortality in clinical settings [30, 31].

TABLE VII. TOP STRUCTURAL FEATURES THAT ARE ASSOCIATED WITH INCREASED 30-DAY MORTALITY RISK BY L2-REGULARIZED LOGISTIC REGRESSION, USING COMBINED FEATURES

| Structural features | Coefficients |
|---|---|
| Metastatic cancer | 0.923 |
| Admission | 0.533 |
| Mechanical ventilation | 0.378 |
| Blood albumin | -0.342 |
| Ph | 0.312 |
| Hemoglobin | -0.204 |
| Blood serum creatinine | -0.158 |
| Magnesium | 0.155 |
| SIRS | 0.149 |
| Temperature | -0.143 |

Figure 1 also shows the ranking of the top 50 bag-of-words features with its coefficients in the same model, in which the font sizes are proportional to the coefficients in the model. Many of the selected features show clinically meaningful words. For instance, in bag-of-words, 'arrest', which had the highest coefficient, could mean that the patients had previously experienced cardiac arrest. This could have occurred in or out of the hospital, but either case could lead to a negative survival outcome [32, 33]. The word 'hemorrhage' could also indicate a representative form extracted from 'subarachnoid hemorrhage' or 'gastrointestinal hemorrhage' or a hemorrhage in other organs. It is also a meaningful word because many studies identified a relationship between hemorrhage and mortality risk [34, 35]. The feature also included the word 'metastatic', which is a similar result to the previous top structural features. The words 'fibrosis', 'PICC' (Peripherally inserted central catheter), 'ascites' are often used regarding critically-ill patients could indicate a higher mortality risk [36, 37].

To handle the missing physiological variables, we used the Multivariate Imputation by Chained Equations (MICE) method. Since MICE method assumes the missing-at-random pattern of the value, it might not match the exact physiological value of patients in real clinical practice, as clinicians often order tests with certain expectations about the likely results. However, many studies have shown that MICE could still be an effective imputation and baseline method compared to other imputation methods because of its simple implementation [38, 39].

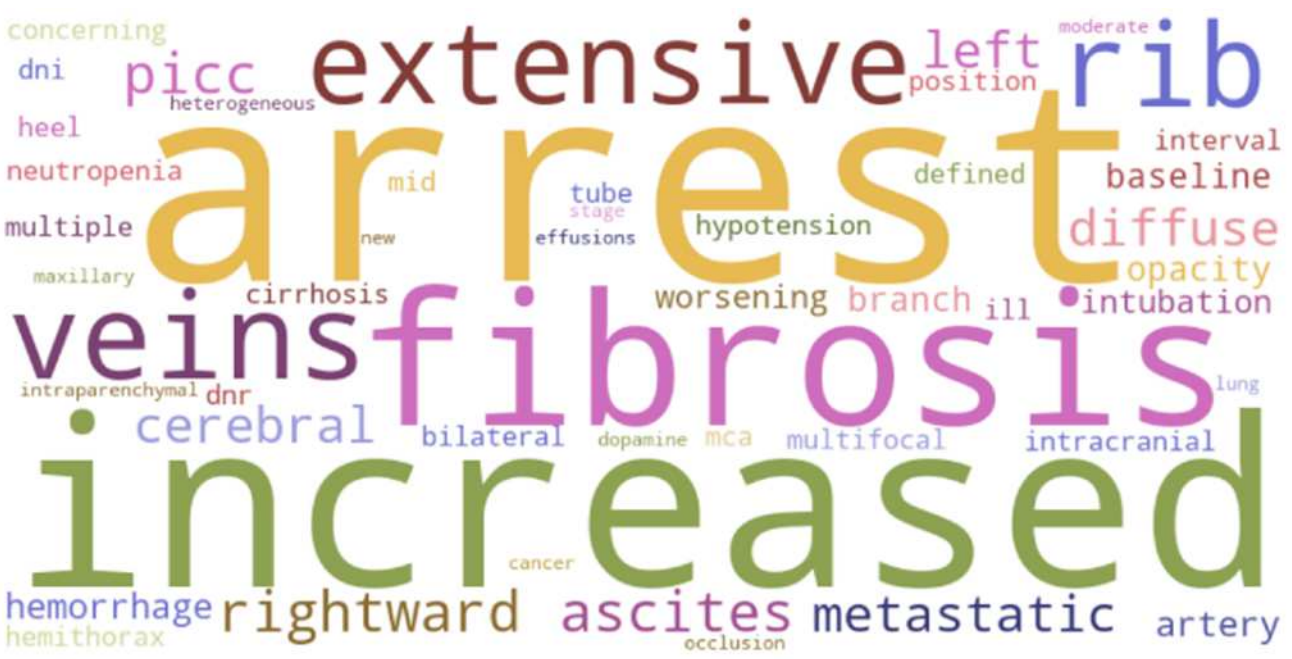

Fig. 1. Ranked top 50 positive features in Bag-of-words with its coefficients as font size in 30-day mortality prediction by L2-regularized logistic regression, using combined features

When working with the clinical note data, we explored the bag-of-words method in conventional machine learning models and the word-embedding in the CNN model. These models showed moderate performance, but the performance of the CNN model was lower than that of the other models. CNN-based architectures generally work well for datasets with short texts, but they may not outperform bag-of-words model on corpus with long texts, such as the clinical note corpus. Considering that training a CNN model is time-consuming, it could be more appropriate to utilize the well-calibrated non-CNN classifier to predict the mortality in our case. There are other options for using medical concepts as features such as through NLP pipelines including MetaMap [40] and clinical Text Analysis. Moreover, there are also other deep learning architectures for handling the contents of long texts. Future investigation could be conducted using those architectures to identify which method is more suitable. Improved model performance could imply the potential application in the clinical setting in the near future. However, more researches and discussions about the model performance are needed to integrate the prediction model to the real-world decision-making in the clinical setting.

In conclusion, our study identified that integrating structural features including demographic/physiological factors and unstructured clinical notes could help to improve the prediction of mortality in adult sepsis patients by using supervised machine learning method. We identified that carefully selected demographic/physiological factors and well-represented clinical notes could predict the mortality in sepsis patients with an AUC greater than 0.84. Our study suggested that the future prospective cohort studies are needed to validate whether our approach is effective to predict the mortality of the sepsis patients at the early in their hospital and initiate the intervention to reduce the mortality risk.